# A Statistical Approach to the Concept of Mass

Martin Tamm

**Abstract**

In this paper, we use methods from differential geometry and statistical mechanics to investigate a model for the concept of mass. The theory is not quantum mechanical in the usual sense, although certain features like multiple histories and uncertainty are included in a natural way.

En utilisant des méthodes de mécanique statistique et de géométrie différentielle, nous étudions un modèle pour le concept de masse. Il ne s'agit pas d'une théorie quantique au sens habituel du terme, même si certaines propriétés, comme la multiplicité des histoires et la notion d'incertitude, sont incluses d'une façon naturelle.

# 1. Introduction.

The purpose of this paper is to try to understand the origin of gravitation and mass. By necessity, any such attempt must involve both the general theory of relativity and quantum theory. Many authors have tried to develop theories of gravity in a way similar to the very successful theories describing other forces in nature, thus in some sense considering quantum physics to be the fundamental theory from which general relativity should be derived. This is, of course, very natural in view of the success of quantum mechanics but, on the other hand, there seems to be no obvious reason why the essentially linear theory developed in the first half of the 20th century should be the logical foundation for the nonlinear theory of gravitation.

The perspective in the present approach is the reversed. In fact, it must be considered as a historical coincidence that the now classical theory of quantum mechanics was developed at a time when general relativity was not yet strong enough to influence it to any considerable effect. Hence, it seems interesting to ask how all the new experimental facts that were discovered in the early history of quantum physics, and that puzzled the scientists of those days so much, would have been interpreted by a scientific community which had already become more familiar with general relativity.

Our starting point is the following: On the one hand, general relativity gives an almost perfect understanding of how matter influences geometry on the macroscopic level. On the other hand, we do not know how this influence is actually created. Thus, in view of the very close link between matter and geometry, it seems natural to ask if this connection could also be used in the reverse direction, i.e. could we use geometry to understand what matter is?

In this paper, we will assume that the presence of a particle somehow implies a topological obstruction to flat geometry. Together with the stochastic nature of quantum physics this will make it possible to give a definition of mass in terms of metric properties like curvature, and investigate how these concepts are related.

The purpose is not to advocate any particular view of what an elementary particle is and no attempt will be made to investigate to what extent these ideas are compatible with other theories that can explain other parts of physics. Rather, the ambition has been to construct a simple enough framework for the mysterious concept of mass to become comprehensible.

The basic tools to be used come from statistical mechanics. Although these methods were developed for other purposes, they still contain our best knowledge of how stochastic microscopic laws lead to a deterministic macroscopic world.

Needless to say, it is by no means obvious that such an approach will lead to better theory than what we already have. But it is not unlikely that its shortcomings will be of quite a different kind, as compared to what we are used to; hence it may contain elements that can contribute to a final understanding of the foundations of physics.

The basic outline of this paper is as follows: In Section 2 we state the main postulates underlying this work. These postulates are not to be considered as a logical foundation from which the rest of the theory should be deduced. Rather, they describe the philosophical framework which was the starting point for this work. In Section 3, we then try to find a natural Ensemble for space-time geometry, i.e. we try to attribute a certain statistical weight to every metric. In Sections 4-6, we use this Ensemble to investigate the most probable geometries. In particular, we will arrive at a purely geometric definition of action and, connected with it, a "Principle of Least Action", which turns out to be the analogue in this context of minimizing the free energy in statistical mechanics. The final three sections are then devoted to showing how this principle is connected to physics as it is known to us. In Section 7, we show that Einstein's field equations follow from the Principle of Least Action. In Section 8, we then introduce a definition of inertial mass which together with the Principle of Least Action result in the usual conservation law for energy. In this sense, the framework we have introduced explains both the fundamental aspects of the concept of mass: the gravitational and inertial aspects.

The geometric perspective on mass which we have adopted in this paper results in one unusual property of this theory: It may be that the strength of the gravitational field generated by a particle (as manifested in the mass-parameter in the corresponding Schwarzschild metric), need not be proportional to its inertial mass. This should in no way be confused with the equivalence between gravitational and inertial mass implied by general relativity, which is a corner stone in the theory. Rather, it concerns the question of *how* and *why* mass generates curvature, a question about which general relativity has little to say.

Although the model in Section 9 is too coarse to give reliable numerical results, the conclusion is that light particles, like neutrinos, could contribute much more to the gravitational field than what usually believed. Thus, this could be related to the missing dark matter of the universe.

## 2. The Basic Postulates.

General relativity can be said to rest on just one extremely natural idea, namely the equivalence of all frames of reference. Although general relativity is a macroscopic theory, we will assume this to hold true also on the microlevel. This leads to the following

**Postulate 1 (Relativity Principle).** Physics in space-time can be described by manifolds which are locally lorentzian, i.e. at every point, the geometry is given by locally equivalent Lorentz frames.

The history of quantum physics on the other hand, is a long list of odd solutions to various counterintuitive problems which were all in the end united into the remarkably efficient theory of quantum mechanics. If we take a closer look at the development of the new theory, there seem to have been mainly four types of phenomena that were involved in the invention of quantum mechanics: First, certain variables classically believed to behave continuously turned out to be discrete. Second, the microscopic world seemed to behave in a nondeterministic way in certain situations and third, it was realized that not all dynamic variables of a particle could simultaneously be given a precise meaning. Finally, in some cases it seemed as if particles could simultaneously develop in different ways and hence many different histories of the same particle had to be taken into account to get a full description of an event.

In this paper, it is the last type of phenomena which will be central. In fact, we will make no attempt at including the whole of quantum mechanics.

This leads to the following

**Postulate 2 (Weak Quantum Principle).** To describe the physics of a certain region in space-time, we must take into account all space-time manifolds compatible with the outer constraints.

Clearly, any attempt at going further must necessarily make this statement in some way more precise. In this paper, we shall, using a well-known idea from statistical mechanics, treat all space-time geometries as an Ensemble, thus attributing a certain statistical weight to each geometry. This will be explained in the next section.

It should be noted that at this stage we make no assumption about the way in which different manifolds interfere with each other, as a more complete theory would clearly have to do. In particular, no reference will be made to the complex phase which plays such a mysterious role in quantum mechanics.

In deciding how this Ensemble should be constructed, we will be guided by the following geometric idea which might, on the one hand, not obviously be true, but on the other hand originates from Einstein's theory of gravitation in a very natural way.

**Postulate 3 (Local Geometric Principle).** The statistical weight of a given space-time manifold only depends on its local geometric properties.

As a consequence, the weights of disjoint regions will be treated as independent variables. More generally, if a region $\Omega$, is split into smaller disjoint sets $D_a$, $a \in I$, with probability weights $\Psi_a$, then the total weight for $\Omega$ will be given by

$$\Psi = \prod_{a \in I} \Psi_a . \tag{1}$$

*Remark* 1. It may be said that this principle could somehow be in conflict with our common views of physics, since quantum mechanics in a certain sense is obviously not local. Since the emphasis in this paper is on mass and gravitation, and not on quantum mechanics, we will not try to argue against (or in favor of) such a standpoint. However, it should be noted that our postulates can give rise to non-local phenomena (see further Remark 12 in Section 6).

Although these three postulates constitute the cornerstones of this approach, we will now also add a fourth which formalizes some ideas, naturally emerging from quantum physics, about the discrete micro structure of space-time itself. Even if there is nothing about them which is obviously true, we consider them to be central enough in modern physics to serve as a starting point for further investigations. However, it seems desirable to find a formulation which is as weak as possible:

**Postulate 4 (Discreteness assumption).** Space-time is somehow constructed from discrete units, to which we will refer as *elementary cells*, the sizes of which are comparable to the Planck length : $10^{-33}$ cm. Moreover, the uncertainty of a measurement of length at this length scale is comparable to the length itself.

There are obvious difficulties connected with trying to apply differential geometry to the kind of universe implied by Postulate 4. These problems would be likely to disappear, in one way or another, if we had a good knowledge about the structure of space-time at very short distances. At the present stage of development, we must simply make use of what we have. Usually, we will simply choose to work on a coarser scale than the Planck scale, where concepts from ordinary differential geometry still make good sense.

Naturally, it can not be expected it to be possible to deduce all laws of physics from these postulates alone in a strict logical sense. However, in the next section, we will start to look for a natural candidate for an Ensemble on the basis of these assumptions.

## 3. The Ensemble.

For the basic Ensemble formalism of this section, we refer to [1] and [3]. However, before proceeding, we will now recall a few basic concepts from statistical mechanics.

Given a classical thermodynamic system with $N$ particles $q_1,...,q_N$, statistical mechanics postulates under very general circumstances the probability for the occurrence of a certain configuration. Possibly the most common way of doing this is called *the canonical Ensemble*, which assigns to a given configuration a probability $p$ proportional to

$$\exp\{-bH(q_1,...,q_N)\}, \qquad (2)$$

where $b$ is inversely proportional to the temperature and $H(q_1,...,q_N)$ denotes the total energy of the system. For this to make sense, it is of course necessary that the sum (or integral, depending on the context)

$$\Pi = \sum \exp\{-bH(q_1,...,q_N)\}, \qquad (3)$$

over all possible states of the system, converges. $\Pi$ is called the *state sum* of the system and is used for many computational purposes in statistical mechanics. The canonical Ensemble is used for systems with a fixed number of particles.

In the following, we will try to use the same approach to assign probabilities to different space-time geometries, with some important differences, however. First of all, the role of particles in the classical theory is now played by the volume elements $D_a$. Since energy is not considered to be a fundamental concept, we will have to ask ourselves what should be used in formulas (2) and (3) above instead of the function $H$. (In fact, one of the purposes of this paper is to arrive at a definition of mass-energy based on more fundamental geometric concepts.)

But the situation is further complicated by the fact that the volume may also vary. This corresponds to the situation in classical statistical mechanics where the number of particles is not fixed, and the standard way of handling this problem is to replace the canonical Ensemble by the so-called *grand canonical Ensemble* which is in some sense what will also be done in Postulate 5 below. Since we do not want to use more terminology from statistical mechanics than necessary, we will simply refer to the resulting probability distribution as the Ensemble.

If we now try to determine the Ensemble, let us first note that, in view of the relativity principle and the local geometric principle, it is a natural starting point to assume that the probability weight of a certain geometry should be a function of the scalar curvature $R$, since this is really the simplest local geometric invariant there is. It might very well be that more complicated invariants are involved but even if so, it is natural to start by investigating the simplest situation and introducing more complicated assumptions only when they are called for. We also note that this type of argument is commonly used to motivate the Hilbert action principle in the theory of general relativity [2]. Since the curvature fluctuations can be expected to become increasingly larger at short distances, it may be meaningless to speak about curvature at points. Therefore, we will only consider mean curvature over sets with a given extension, and assume that this to contain all physical information about curvature that we need. This means curvature will still be denoted by $R$. Clearly, the probability weight will also have to depend on the volume of the region if formula (1) is to make any sense.

Consider a region $\Omega$ in space-time, and suppose that we write $\Omega$ as a union of sets $D_a, a \in I$, all with approximately equal space-time diameter ($\approx d$) and volume ($\approx d^4$). These sets will be considered as small compared to the size of elementary particles but still large compared to the Planck length. The volume of each $D_a$ will fluctuate in the following, but whenever we have a partition $\Omega = \bigcup D_a$ of this kind, we will assume each $D_a$ to consist of a fixed number of elementary cells as in Postulate 4. Moreover, we will only be interested in situations where the underlying metric varies slowly on each $D_a$, and not in the detailed behaviour of the metric fluctuations inside each of the small regions, $D_a$.

Let us start by investigating the curvature-dependence to see what kind of dependence could be expected. It is reasonable to argue that $R_a$, the mean curvature in $D_a$, is the sum of contributions from fluctuations of the metric in much smaller regions, and that these fluctuations behave, roughly, as statistically independent variables. Hence, if we in addition assume the expectation value to be zero, a central limit type of argument shows that the probability amplitude for $R_a$ should behave as

$$P_a \propto \exp\{-b_d R_a^2\}. \tag{4}$$

Another way of obtaining the same formula, familiar from statistical mechanics, is the following: Since the $D_a$:s are large as compared to the Planck length, it is reasonable to assume that only small values of $R_a$ will play a significant role. Hence, if we let $y_a = -\log \Psi_a$, then only the lowest terms in the Taylor expansion of $y_a$ around $0$ need to be considered. If, in addition, we assume $y_a$ to be symmetric in $R_a$, we can write

$$y_a \approx y_0 + b_d R_a^2. \tag{5}$$

If we use that $\Psi_a = \exp\{-y_a\}$ and note that the factor $\exp\{-y_0\}$ only contributes with a common multiplicative factor, which does not influence the probabilities, then we once more arrive at (4).

If we accept this formula for $\Psi_a$, the statistical weight of the geometry within $D_a$, then in view of the local geometric principle, we are led to the following expression for the joint probability weight:

$$\Psi = \prod \Psi_a = \exp\{-b_d \sum_a R_a^2\}. \tag{6}$$

The parameter $d$ is our own choice, and we do not expect it to have any physical significance in itself in this context. Therefore, it is important to note how our conclusions depend on $d$ and, in the end, we hope all such influence to cancel out from formulas with physical implication. For instance, it follows from the multiplicative property (1) that the parameter $b_d$ must be of the form

$$b_d = d^4 b, \tag{7}$$

for some constant $b$ independent of $d$. In view of this, (6) can therefore approximately be

rewritten as
$$\Psi = \exp\{-b \int_\Omega R^2 \, dV\}, \tag{8}$$
where $dV$ is the four-dimensional volume element.

If we now turn to the volume fluctuations, then just as for curvature, we expect these to be roughly described by a normal distribution. However, the situation is in this case complicated by the influence that the $D_a$:s may exert on each other. Once more, we consider the partition $\Omega = \mathbf{U} D_a$, and assume the probability weight for each $D_a$ as a function of volume to be given by
$$\Psi_a = \Psi(V_a), \tag{9}$$
where $V_a$ is the volume of $D_a$. In free space, we expect there to be a certain "equilibrium volume" $V_0 \approx d^4$ around which $V_a$ oscillates. Since we suppose the $D_a$:s to be large in comparison with the fluctuations, we can write:
$$y_a = -\log \Psi_a = y_a(0) + y'_a(0)(V_a - V_0) + \frac{1}{2} y''_a(0)(V_a - V_0)^2 + \ldots \tag{10}$$
This can be more conveniently expressed by introducing the quantity
$$r_a = \frac{V_a - V_0}{V_0}. \tag{11}$$
$r_a$ clearly measures the relative deviation of the volume from its equilibrium value in each $D_a$. Although this name might be somewhat misleading, we will refer to $r$ as the *density*.

If we now write $p_0 = y'_a(0)$ and omit the terms independent of $V_a$ in (10), we get
$$y_a = p_0 V_a + k_d r_a^2, \tag{12}$$
and
$$\Psi_a = \exp\{-p_0 V_a - k_d r_a^2\}, \tag{13}$$
where $k_d$ is a constant. Taking the product as before, we obtain
$$\Psi = \prod \Psi_a = \prod \exp\{-p_0 V_a - k_d r_a^2\}, \tag{14}$$
and as in (7), the dependence on $d$ must be of the form
$$k_d = d^4 k, \tag{15}$$
for some fixed constant $k$ independent of $d$. Once more, (14) can be rewritten as
$$\Psi \approx \exp\{-p_0 V(\Omega) - k \int_\Omega r^2 \, dV\}. \tag{16}$$

*Remark* 2. In classical statistical mechanics, the pressure can be computed (up to a factor $1/b$) as the volume derivative of the logarithm of the state sum. If we for a moment assume the state sum to be dominated by a state with $r = 0$, we see that the constant
$$p_0 = -\frac{\Delta \log \Psi}{\Delta V} \tag{17}$$
can, in analogy with the classical case, be interpreted as (the negative of) the "geometric pressure". In this paper, we will, for stability reasons, assume that $p_0 > 0$. It is possible that the opposite assumption could also lead to a stable theory, but we will not deal further with that issue here.

Moreover, we can get a more general formula, valid also for small but non-zero $r$, by noting from (12) that

$$p = -\frac{\Delta \log \Psi}{\Delta V}\left(\approx \frac{d}{dV_a}\left(p_0 V_a + k_d r_a^2\right)\right) = p_0 + 2kr, \tag{18}$$

where we have used (11) and (15). $p$ is thus locally computed and can hence vary from point to point.

If we now make the assumption that curvature and volume can be treated as independent, the above expressions in (8) and (16) can be multiplied together to give

**Postulate 5 (Ensemble Postulate).** Given a bounded region $\Omega$ as above, and assuming that the geometry is specified on $\partial \Omega$, the weight of a certain space-time geometry in the interior, specified by a metric and a density $(g, r)$, and compatible with the boundary conditions, can be written as

$$\Psi = \exp\{-p_0 V(\Omega) - k\int_\Omega r^2\, dV - b\int_\Omega R^2\, dV\}. \tag{19}$$

Of particular importance is the case of constant volume:

$$\Psi = \exp\{-k\int_\Omega r^2\, dV - b\int_\Omega R^2\, dV\}. \tag{20}$$

*Remark* 3. The assumption that curvature and volume can be treated as independent is not an obvious one. Although there may be no sense in asking whether this is true or false without any knowledge about what actually determines the geometry at the Planck scale, it can easily be imagined that, say, diminished volume and negative curvature could be correlated. Still, in this preliminary investigation we have chosen to make this assumption for two reasons: First, it considerably simplifies the calculation. Second, it does not appear to have any considerable influence on the general conclusions. However, it may be worthwhile to pause here to see how Postulate 1 would look under more general assumptions.

Thus, we return to (9) but now assume that $\Psi$ simultaneously depends on both $V_a$ and $R_a$. Expanding $y$, we get:

$$y = y_0 + p_0(V_a - V_0) + q_0 R_a + k_d r_a^2 + b_d R_a^2 + h_d r_a R_a + \mathbf{L}, \tag{21}$$

where the terms $y_0$ and $q_0 R_a$ can be dropped as before. Continuing the argument as above and rewriting in integral form, we obtain instead of (19)

$$\Psi = \exp\{-p_0 V(\Omega) - \int_\Omega Q(r, R)\, dV\}, \tag{22}$$

where $Q$ is the quadratic form

$$Q = kr^2 + bR^2 + hrR. \tag{23}$$

It is important for the following that $Q$ is strictly positive definite. On the other hand, as long as this condition is fulfilled, the exact value of $h$ is of less importance.

*Remark* 4. We have chosen to use integrals in the formulation of the postulate, since this way of writing appears to be simpler and more suggestive. However, it should be remembered that the meaning and existence of a well-defined concept of volume at very short distances is unclear, and that the integral notation here is only a way of rewriting the previous summation formulas. In fact, in view of Postulate 4, the Ensemble should be thought of as being essentially discrete.

*Remark* 5. As mentioned at the beginning of this section, for this approach to make sense it is necessary to know that the "state-sum", i.e. the sum of all unnormalized probabilities of the Ensemble,

$$\Pi = \int \exp\{-k \int_D r^2 \, dV - b \int_D R^2 \, dV\}, \tag{24}$$

converges, since otherwise there is no way of normalizing the Ensemble to get purposeful probabilities. To actually prove this mathematically, however, turns out to be an enormously complicated task. Not only do we have to include all possible metrics on a given manifold in the summation/integration above, but we do, in fact, also want to allow for different topological structures, which leads to still greater problems.

This may be quite unsatisfactory from the point of view of mathematics. However, for a theory of physics, it may be less important. In fact, what we actually want to do with the Ensemble in the following sections is to compare the probabilities of different metrics on a scale which is large as compared to the sets $D_a$. For this purpose, we need only compare the corresponding sums over metrics which are so close to the given ones that they can be treated as perturbations. In this case, it is comparatively easy to give a reasonable interpretation of the integral in (24) above. In this paper, this can be seen in the computations leading to formulas (61) and (67) in Section 6.

## 4. Macro states.

The purpose of this section is to try to find the most probable geometry on a scale which is large compared to the quantum fluctuations. This is analogous to the central task in statistical mechanics which consists in deducing thermodynamical properties from the Ensemble. Note that "large" in this context may still refer to sub-atomic distances.

Before proceeding to this problem, however, we must comment on the use of the word "geometry" here. It is an important and beautiful property of general relativity that all the geometry you need is contained in the metric. In the previous section, we have, in addition to the metric, introduced the density $r$ as an additional variable. In fact, if we for instance consider a region $\Omega$ in space with the flat Minkowski metric, then two geometries in $\Omega$ with this metric but different (constant) values of $r$, must be considered to be different. Thus, it could appear that $r$ has a significance, independent from the metric. This would, however, be a false impression, since the definition of $r$ itself in (11) is based on the metric. Rather, the introduction of $r$ should be considered as the necessary (but well worth) price that must be paid if we want to approximate the discrete space-time of Postulate 4 with differentiable manifolds, where classical differential geometry applies. In a more fundamental approach, based on a discrete metric, $r$ would be superfluous.

In the following, we will denote a geometry with metric $g$ and density $r$ by $G = (g, r)$ or $G = (g, r_G)$ if necessary to avoid ambiguity.

So far, we have only considered a bounded region $\Omega$ in space-time, where the geometry has been given on $\partial\Omega$. We can now generalize this concept to arbitrary space-time manifolds. First, observe that instead of considering the geometry to be fixed on $\partial\Omega$, we might as well more generally assume that we are given any probability measure on the set of geometries on $\partial\Omega$, and then study the corresponding average Ensemble in $\Omega$.

Now, assume that we are given two bounded regions $\Omega$ and $\Omega'$ in space-time with $\Omega \subset \Omega'$ and, furthermore, an Ensemble measure $dP_{\Omega'}$ on $\Omega'$. Then, clearly $dP_{\Omega'}$ can be used to define a new Ensemble measure $dP_{\Omega}$ on $\Omega$ by assigning to each set $U$ of geometries on $\Omega$, the average measure with respect to $dP_{\Omega'}$ of all geometries in $\Omega'$, such that their restrictions to $\Omega$ belong to $U$. We say that $dP_{\Omega}$ is the Ensemble measure on $\Omega$ induced by $dP_{\Omega'}$.

**Definition 1.** A *Space-time state W* is a topological 4-manifold $M$ with an Ensemble measure $dP_{\Omega}$ for each bounded subset $\Omega$, such that whenever $\Omega \subset \Omega'$, $dP_{\Omega}$ is induced by $dP'_{\Omega}$ as above.

Clearly, every compact manifold-with-boundary $\Omega$ with an Ensemble measure $dP_{\Omega}$ will give rise to a space-time state by equipping each subset $\Omega'$ with the corresponding Ensemble measures induced by $dP_{\Omega}$.

On the other hand, in the case of a non-compact manifold, we may attempt to construct space-time states by taking suitable limits of Ensemble measures on bounded subsets. This should be compared with the construction of Gibbs measures in Statistical Mechanics (see [4]), although the difficulties involved would be much greater due to the lack of translation invariance.

Given as space-time state, we will, as is common in statistical mechanics, assume that the partition sum $\Pi$ for a certain region $\Omega$ is dominated by one or a few metrics at most, or rather, by the set of small perturbations to these metrics:

**Definition 2.** Given a space-time state $W$, a *macro geometry* for $W$ is a geometry which (together with all small perturbations of it) maximizes the probability among all other geometries and for all bounded sets, $\Omega$.

In the following, we shall consider the main problem of physics (in our limited context) to find the macro-geometries of a given space-time state.

*Remark* 6. In this definition, we have tacitly assumed that the geometries that we compare, only change on a much larger scale than the Planck scale and thus can be separated from the perturbations. In fact, a metric which dominates $\Pi$ is expected to have strong regularity properties. To prove this mathematically may be very difficult but the reasons are partly obvious: If a possible singularity manifests itself through high scalar curvature, then clearly the exponentiation in Postulate 5 will guarantee that it will play no important role in the partition sum. On the other hand, manifolds with zero scalar curvature can be quite singular. But it will be argued in Section 6 that not only large scalar curvature but in fact also large Ricci tensor makes the corresponding metric improbable, which puts much more severe conditions on the most probable metrics. Thus only very regular geometries can be candidates for maximizing the probability.

It should be noted, however, that these regular geometries can still be extremely singular from the point of view of usual general relativity. Typically, we shall assume that changes in $g$ may at most take place on a scale characteristic to elementary particles, whereas the fluctuations become dominant when we approach the Planck length.

To compute the probability of such a macro geometry we shall, in view of the last remark, treat the curvature and density of the underlying geometry $G = (g, r_G)$, and the fluctuations $(dg, dr)$, as statistically uncorrelated. In particular, for the curvature this means that we neglect the cross-term

$$2\int_\Omega R_g \, dR \, dV, \tag{25}$$

to get an approximate equality

$$\int_\Omega R^2 \, dV = \int_\Omega (R_g + dR)^2 \, dV \approx \int_\Omega (R_g^2 + dR^2) \, dV, \tag{26}$$

which will hold true with overwhelming probability if the set $\Omega$ is large in comparison with the length $d$ (the diameter of the sets $D_a$).

In a completely similar manner, we can also write

$$\int_\Omega r^2 \, dV \approx \int_\Omega (r_G^2 + dr^2) \, dV. \tag{27}$$

To each macro-geometry $G$, we now associate an (unnormalized) probability as follows:

$$\Pi_G \approx \int \exp\{-p_0 V(\Omega) - k \int_\Omega r^2 \, dV - b \int_\Omega R^2 \, dV\}. \tag{28}$$

Here, the integration is carried out over all perturbations of the given metric. This means integration over all perturbations $dg_a^{ij}$ of the metric coefficients in each set $D_a$, which leads to enormously complicated integrals. In practice, however, this will be drastically simplified by using a kind of average approximation (see (62) below).

$$= \int \exp\{-p_0 V(\Omega) - k \int_\Omega (r_G^2 + dr^2) \, dV - b \int_\Omega (R_g^2 + dR^2) \, dV\} \tag{29}$$

$$\approx \exp\{-p_0 V(\Omega) - k \int_\Omega r_G^2 \, dV - b \int_\Omega R_g^2 \, dV\} w_G(\Omega), \tag{30}$$

where

$$w_G(\Omega) = \int \exp\{-k \int_\Omega dr^2 \, dV - b \int_\Omega dR^2 \, dV\}. \tag{31}$$

Since volume and curvature fluctuations are treated as independent according to the discussion in Section 3, it follows that $w_G$ splits into two parts,

$$w_G(\Omega) = \int \exp\{-k \int_\Omega dr^2 \, dV\} \int \exp\{-b \int_\Omega dR^2 \, dV\}. \tag{32}$$

The first factor in this formula is just a constant which is unimportant for the probabilities; hence, it will be dropped in the following. As for the second factor, it is independent of the density, but depends on the metric and will hence from now on be denoted by $w_g$. To compute $w_g(\Omega)$ for a given region $\Omega$ is difficult, and we will return to this problem in Section 6.

We can now, in analogy with the classical theory, introduce the following function of the metric:

$$\Xi_G = -\log \Pi_G, \tag{33}$$

which can also, using (30), be expressed as

$$\Xi_G = k \int_\Omega r_G^2 \, dV + b \int_\Omega R_g^2 \, dV - \log w_g(\Omega) + p_0 V(\Omega). \tag{34}$$

This function is, in a certain sense, analogous to the free energy in statistical mechanics (see

[1] or [3]). For reasons that will become clear below, we will in the following refer to it as the *geometric action* of the macro geometry. If we consider independent fluctuations in the different cells $D_a$, then the term $\log w_g(\Omega)$ is essentially locally determined, and hence, it is natural to introduce the corresponding density $m_g$, i.e. $m_g$ is defined by the requirement that

$$\log w_g(D) = \int_D m_g \, dV, \tag{35}$$

for every $D$. $m_g$ may thus be considered of as a kind of "effective (logarithmic) density of states".

We can now rewrite (34) as

$$\Xi_G = \int_\Omega \left( kr^2 + bR_g^2 + m_g \right) dV + p_0 V(\Omega). \tag{36}$$

In particular, if we consider a region $\Omega$ of constant volume, the last term is just a constant that can be omitted, and we then write

$$\Xi_G = \int_\Omega \left( kr^2 + bR_g^2 + m_g \right) dV. \tag{37}$$

In classical statistical mechanics, the properties of the Ensemble are determined by an interplay between the probability of a certain state (depending on the energy) and the number of states with similar marcoscopic properties (the density of states).

Similarly, in our context, the properties of the Ensemble are determined by an interplay between the three terms in (37).

*Remark* 7. When computing the change in $\Xi_G$ due to changes in the volume if $\Omega$, we can omit the curvature dependent terms $R_g^2$ and $m_g$ from (37). Since curvature and volume are treated as independent, it follows from (36), in the case $r = 0$, that

$$\frac{\Delta \Xi}{\Delta V} = p_0, \tag{38}$$

and more generally, as in Remark (2), that (locally)

$$\frac{\Delta \Xi}{\Delta V} = p = p_0 + 2kr. \tag{39}$$

Thus, once more we see that $p_0$ and $p$ can be interpreted as pressure.

*Remark* 8. To only consider fluctuations at the $d$-scale is of course a simplification, and could at most be expected to give a kind of first order approximation. In fact, it is reasonable to expect fluctuations on all length scales. We will return to comment on this in Remark 12 in Section 6. However, from the point of view of gravitation, the possible extension of the approach indicated there would not have any essential influence on the conclusions of the theory. Therefore, in this preliminary investigation, we will concentrate on the simpler version of the theory.

The action integral (37) gives a measure of the probability weight of a given macro geometry. Hence, we can now formulate the following principle which is, in a certain sense, analogous to the principle of minimal free energy in statistical mechanics:

**Principle 1 (of Least Action).** *If the volume of $\Omega$ is given, then the macro geometries are obtained by minimizing*

$$\Xi_G = \int_\Omega \left(kr^2 + bR_g^2 + m_g\right)dV. \tag{40}$$

Making use of the elementary principle of Calculus which states that the derivative of a function at a minimum must be zero, as a particular consequence of Principle 1, we get the following

**Claim 1.** *Consider a macro geometry in $\Omega$ and a sufficiently differentiable volume-preserving infinitesimal transformation $\Phi$ of $\Omega$ which leave $\partial\Omega$ fixed. Then $\Phi$ does not change the action of the macro geometry.*

*Remark* 9. Naturally, the symbols $b, k$ and $m$ have, of course, been chosen with a certain analogy to statistical mechanics in mind. In fact, it will be natural to refer to $b$ as the "the geometric temperature". $k$, on the other hand, is somewhat analogous to the classical compressibility, and will thus be called "the geometric compressibility". $m$ is perhaps less analogous to the classical theory, but it will still be referred to as "the geometric potential".

Note also that $b$ and $k$ are constants given once and for all, whereas $m_g$ depends on the given geometry. It is an important part of this approach to find methods to compute $m_g$ for a given metric.

In practice, when applying the Principle of Least Action, we do want to compare geometries which assign different volumes to $\Omega$. In this case, we may assume $\Omega$ to be contained in a much larger set $\Omega'$, which can serve as a classical "heat-reservoir". Then, we may still find the most probable states in $\Omega$ by minimizing the action integral (40), but now under the restriction that we only consider states in "equilibrium" with the exterior. Once more reasoning in analogy with the classical case, we arrive at the following

**Principle 2 (of Least Action).** *If the volume of $\Omega$ is not assumed to be fixed, the macro geometries are obtained by minimizing*
$$\Xi_G = \int_\Omega \left(kr^2 + bR_g^2 + m_g\right)dV, \tag{41}$$
*among all geometries exerting the same (geometric) pressure on the exterior.*

To make this condition of equilibrium somewhat more precise, for simplicity consider the case of a cylinder $\Omega = \mathrm{i} \times A$ in Minkowski space with $r=0$, where $\partial A$ is a three-dimensional sphere with radius $s$, but where the interior of $\Omega$ has been replaced by some more general geometry, which we suppose to be time-independent, however. If we now consider an infinitesimal transformation $\Phi$ which moves the boundary of $\Omega$ the distance $\Delta s$ outwards, then the macro geometry inside $\Omega$ with action $\Xi$ (/unit of time) will, as a result of this dilatation, be replaced by a new macro geometry with action equal to $\Xi + \Delta\Xi$ (/unit of time). Clearly,
$$P = \frac{\Delta\Xi}{\Delta s} \tag{42}$$
can be interpreted as a kind of total geometric pressure (/unit of time) on the cylinder, and for any geometry in $\Omega$ which is to compete in the Principle of Least Action, this pressure must equal the corresponding pressure from the outside which, in turn, must clearly equal the pressure of a similar cylinder in flat Minkowski space, if the geometry in $\Omega$ should be in equilibrium with empty space. We now proceed to compute this pressure with a method and notation that can later (in Section 9) be generalized.

Hence, we take the macro geometry in $\Omega$, characterized by $r = 0$ and the usual flat Minkowski-metric

$$g_0 = -dt^2 + dr^2 + r^2 df^2 + r^2 \sin^2 f dq^2, \qquad (43)$$

and compare it with the state obtained by stretching this metric out to a cylindrical set with radius $s + \Delta s$. Clearly, the modified metric is given by

$$g_1 = -dt^2 + l \left( dr^2 + r^2 df^2 + r^2 \sin^2 f dq^2 \right), \qquad (44)$$

where $l = (s + \Delta s)/s = 1 + \Delta s/s$. By elementary geometric consideration, such an infinitesimal increase in the length scale will result in a three times larger change in the volume scale; thus we get for the density (to first order):

$$\Delta r = \frac{3\Delta s}{s}. \qquad (45)$$

To compute the change in action, we note (from (42) and the definition of $r$ (11)) that

$$\Delta \Xi = p_0 \times \Delta V = p_0 \times V \times \Delta r. \qquad (46)$$

Thus, in our case, we get

$$\Delta \Xi = p_0 \times \frac{4}{3} p s^3 \times \frac{3\Delta s}{s} = p_0 4 p s^2 \Delta s, \qquad (47)$$

or, in other words,

$$P = p_0 4 p s^2. \qquad (48)$$

In particular, the total pressure per unit area of the cylinder is equal to the local $p_0$ as would be expected. It should, however, be noted that the geometric pressure here differs in many other respects from e.g. the pressure of an ideal gas. For instance, it does not necessarily follow that geometric pressure is independent of the form of the boundary of the region.

## 5. Computations with the curvature tensor.

In this section, we shall study variations of the average scalar curvature. Thus let

$$R_D = \frac{1}{V(D)} \int_D R \, dV, \qquad (49)$$

where $D$ is some (small) set. In the following, all variations will have compact support in $D$.

To start with, we compute the first variation of $R_D$. We note that

$$dR_\Delta = d\left( \frac{1}{V(D)} \int_D R \, dV \right) = -\frac{d(V(D))}{V(D)^2} \int_D R \, dV + \frac{1}{V(D)} d\left( \int_D R \, dV \right). \qquad (50)$$

Now, by a well known result (see [2] or [6])

$$d(\int_D R \, dV) = \frac{1}{V(D)} \int_D \left( R_{ij} - \frac{1}{2} g_{ij} R \right) dg^{ij} \, dV \qquad (51)$$

$$+ \frac{1}{V(D)} \int_D g^{ij} (d\Gamma^k_{ij;k} - d\Gamma^k_{ik;j}) dV,$$

where $R_{ij}$ denotes the Ricci tensor and the second integral, by a covariant partial integration, can be shown to vanish when the support of $dg^{ij}$ is contained in $D$. Therefore, for the first variation we get:

$$d(R_D) = -\frac{d(V(D))}{V(D)^2} \int_D R \, dV + \frac{1}{V(D)} \int_D \left( R_{ij} - \frac{1}{2} g_{ij} R \right) d g^{ij} \, dV \tag{52}$$

Hence,

**Claim 2.** *In the case of vanishing Scalar curvature and when the underlying metric g varies slowly compared to the size of D, we get*

$$d(R_D) \approx R_{ij} d g_D^{ij}, \tag{53}$$

*where $d g_D^{ij}$ denotes the averaged variation and $R_{ij}$ is the Ricci tensor of g.*

Furthermore, if the Ricci tensor also vanishes, then the leading terms in the expansion of $R_D$ make up a second order expression in the $d g_a^{ij}$:s. Computing these terms is a task requiring some labour. We will now consider the case of flat space-time and later use this case as approximation of the situation where the underlying metric varies very slowly in comparison with the metric fluctuations. Since, in this case, $R_{ij}$ and its first order variation vanish, for the second variation we obtain

$$d^2 R_D = d^2 \left( \frac{1}{V(D)} \int_D R \, dV \right) = \frac{1}{V(D)} \int_D g^{ij} (-g)^{1/2} d^2 R_{ij} \, dV. \tag{54}$$

To compute $d^2 R_{ij}$, we make use of the well-known formula ([2], [6]):

$$R_{ij} = \sum_k \frac{\partial}{\partial x^k} \Gamma_{ij}^k - \frac{\partial}{\partial x^i} \left( \sum_k \Gamma_{kj}^k \right) + \sum_{k,l} \left( \Gamma_{ij}^l \Gamma_{lk}^k - \Gamma_{kj}^l \Gamma_{li}^k \right). \tag{55}$$

Clearly, since each variation $d g^{ij}$ has compact support in $D$, the same is true for all the $d\Gamma_{ij}^k$:s. This means that integration will cancel out the first terms in (55) which, in turn, means that they will not influence the second variation. Furthermore, since all Christoffel symbols vanish for the flat metric, we easily see that the only still remaining part in $d^2 R_{ij}$ is

$$d^2 R_{ij} = \sum_{k,l} \left( d\Gamma_{ij}^l d\Gamma_{lk}^k - d\Gamma_{kj}^l d\Gamma_{li}^k \right). \tag{56}$$

We now express the $\Gamma_{ij}^k$:s in terms of metric coefficients by

$$\Gamma_{ij}^k = \frac{1}{2} \sum_l g^{kl} \left( \frac{\partial g_{jl}}{\partial x^i} + \frac{\partial g_{il}}{\partial x^j} - \frac{\partial g_{ij}}{\partial x^l} \right), \tag{57}$$

and insert in (55). Making use of the fact that all partial derivatives of the flat metric vanish, we obtain a sum

$$\frac{1}{4V(D)} \sum \pm \int \frac{\partial}{\partial x^k} \left( d g_{ij} \right) \frac{\partial}{\partial x^l} \left( d g_{mn} \right) dV. \tag{58}$$

containing more than 200 terms, most of which vanish due to the compact support of each $d g_{ij}$, however. Clearly, the obtained expression is thus an indefinite quadratic form in the (first derivatives of the) perturbations of the metric.

## 6. The geometric potential.

Let us start by considering the geometric potential in flat space-time and then try to determine the lowest order correction in the presence of curvature.

To compute the geometric potential $m_g$, we must first compute $w_g$. This is not easy however, and we will clearly have to work with some kind of approximations. The following construction is not unique, since it involves certain choices. On the other hand, all the computations to come will be very insensitive to how these choices are made.

Thus, we choose differentiable functions $f_a$ with support in the cells $D_a$. In view of the assumptions in Section 3, this can be done in such a way that they are more or less translates of each other. A variation of the metric $g$ will now mean that we add a real multiple of the corresponding $f_a$ to each of the components $g^{ij}$ in each $D_a$, i.e.

$$dg_a^{ij} = t_a^{ij} f_a \tag{59}$$

where the $t_a^{ij}$:s are numbers, only subject to the obvious condition $t_a^{ij} = t_a^{ji}$. In practise, we will continue to write $dg_a^{ij}$ for these variables, but treat them as one-dimensional. In the following, we will compute the number of states by integrating with respect to these variables:

$$w_g = \int \exp\{-b \int_\Omega dR^2 \, dV\} \tag{60}$$

$$\approx \prod_{a \in I} \int_{\mathbf{R}^{10}} \exp\{-b_d dR_a^2\} dm_a, \tag{61}$$

where $dG = \prod_{a \in I} dm_a$ with $dm_a = \prod_{i \leq j} d(dg_a^{ij})$.

In the case of vanishing Ricci tensor, i.e. when the first-order variation of $R$ is zero, we have in (58) expressed the second-order variation as an indefinite quadratic form in the derivatives of $dg$. Hence, the leading term in the expansion of $R^2$ is a semi-definite form of order 4. Since the exact expression for this form is very complicated, it is not easy to exactly compute the corresponding integral, even in the case of flat space. We shall therefore make a simplifying approximation and replace it by a kind of average perturbation. Thus, we write

$$(\text{"average of"}) \quad (dR)^2 : h\|dg\|^4 \quad \text{where} \quad \|dg\| = \left(\sum_{ij} g_{ij}^2\right)^{\frac{1}{2}}, \tag{62}$$

and $h$ is some constant.

*Remark* 10. Clearly, this approximation may be very poor if we restrict ourselves to a single set $D_a$. On the other hand, the approximation may be fairly reasonable when we simultaneously deal with a large number of variables: Statistically, the variations will be more or less evenly distributed in all directions.

To find a very coarse estimate for the constant $h$, we proceed as follows: First, we observe that it follows from (58) that it is reasonable to write

$$(\text{"average of"}) \quad dR_\Delta^2 = \text{const} \cdot \|\nabla dg\|^4, \tag{63}$$

where the constant is comparable in size to the square of the norm of the quadratic form in (58), i.e. to a first approximation, it is of the order of unity. Using the scaling property $\|\nabla dg\| : d^{-1}\|dg\|$, where $d$ is the diameter of the cells $D_a$, we get the relationship

$$h\|dg\|^4 : dR^2 : d^{-4}\|dg\|^4. \tag{64}$$

Thus, we conclude that $h : d^{-4}$.

With these facts in mind, we can hence write for $w_0$ in a region $D$ containing $N$ cells $D_a$, in the case of vanishing Ricci tensor:

$$w_0 \approx \prod_{a \in I} \int_{\mathbf{i}^{10}} \exp\{-b_d dR_a^2\} dm_a, \tag{65}$$

$$\approx \left( \int_{\mathbf{i}^{10}} \exp\{-b_d h \|dg\|^4\} dm_a \right)^N, \tag{66}$$

and thus for the the geometric potential,

$$m_0 \approx \frac{N}{V} \log \left( \int_{\mathbf{i}^{10}} \exp\{-b_d h u^4\} du \right), \tag{67}$$

where $V = V(D)$ is the volume of $D$.

The effect on the scalar curvature of a small perturbation of the metric depends on the Ricci tensor as indicated by Claim 2. In particular, if the scalar curvature vanishes but the Ricci tensor is different from zero, small variations in $g$ may cause $R$ to change much more rapidly than in the case of flat space-time which, in turn, implies that fewer perturbations will contribute to the $\Pi_G$ in (28).

To make this somewhat more precise, we note that, referring to the discussion in Section 5, the first order variation $dR$ is given by a linear functional on the space of perturbations $dg$ of $g$:

$$dR = R_{ij} dg^{ij}. \tag{68}$$

In the direction of maximal growth (i.e. when $dg^{ij}$ is proportional to $R_{ij}$), we have by (the equality in) Schwarz inequality:

$$|dR| = \|\Re\|_2 \|d\hat{g}\|_2, \tag{69}$$

where $d\hat{g}$ denotes the component of $dg$ in the $R_{ij}$-direction, and we use the notation $\Re$ for the Ricci tensor. In addition, we also have the vacuum fluctuation term (62) which we simply add to the above, hence obtaining the formula

$$(dR)^2 \approx h \|dg\|^4 + \|\Re\|_2^2 \|d\hat{g}\|^2. \tag{70}$$

**Lemma 1.**

$$\int_{\mathbf{i}^{10}} e^{-(u_1^2 + \ldots + u_{10}^2)^2 - t u_1^2} du = \int_{\mathbf{i}^{10}} e^{-(u_1^2 + \ldots + u_{10}^2)^2} du \cdot (1 - gt + O(t^2)),$$

where $g \approx 0.13$.
(Taylor expansion and numerical integration.)

Using this lemma, we now proceed to the geometric potential (where $V$ is so small that we can neglect the change of $g$ in $V$, and $N$ as previously):

$$m_g = -\frac{\log(w)}{|V|}$$

$$= -\frac{1}{V} \log \left( \prod_{a \in I} \int_{\mathbf{i}^{10}} \exp\{-b_d \left( h \|dg\|^4 \right) + \|R_g\|_2^2 \|d\hat{g}\|^2\} dm_a \right)$$

$$= -\frac{1}{V} \sum_{a \in I} \log \left( \int_{\mathbf{i}^{10}} (b_d h)^{-5/2} \exp\{-u^4 - (b_d/h)^{1/2} \|R_g\|_2^2 u^2\} du \right)$$

$$\approx -\frac{1}{V}\sum_{a\in I}\log\left(\int_{i^{10}}(b_d h)^{-5/2}\exp\{-u^4\}du\left(1-g(b_d/h)^{1/2}\|R_{\Delta,a}\|_2^2\right)\right)$$

$$\approx -\frac{N}{V}\log\left(\int_{i^{10}}\exp\{-b_d h u^4\}du\right)+\frac{N}{V}g(b_d/h)^{1/2}\|R_g\|_2^2$$

$$= m_0 + t\|R_g\|_2^2, \tag{71}$$

where $t : (N/V)\times g(b_d/h)^{1/2}$.

*Remark* 11. Let us note that due to the scaling properties $b_d = bd^4$, $h : d^{-4}$ and $N : d^{-4}$, the number $t$ is independent of $d$. Thus, the geometric potential can be computed using any $d$, at least as long as $R_g$ varies slowly on the corresponding length scale.

Summarizing, we get the following formula for the action $\Xi_g$:

$$\Xi_G = \int_\Omega \left(kr^2 + bR_g^2 + m_0 + t\|\Re_g\|^2\right)dV + p_0 V(\Omega). \tag{72}$$

Note that $m_0$ is a constant, independent of the fluctuations; hence, it can be included in $p_0$ to form a kind of "effective pressure" which, for ease of notation, we will continue to denote by $p_0$ in the following. Comparing with Principle 2, we now arrive at the following form of:

**Principle 3 (of Least Action).** *For a geometry $G$ in a bounded region $\Omega$, let the geometric action be defined by*

$$\Xi_G = \int_\Omega \left(kr^2 + bR_g^2 + t\|\Re_g\|^2\right)dV.$$

*The probability maximizing geometries are characterized by the fact that they minimize $\Xi_G$ among all geometries in equilibrium with the exterior of $\Omega$ (as in Principle 2).*

This formulation will be used in Section 9 to compute the strength of the gravitational field.

*Remark* 12. As has already been stated, the splitting $R = R_g + dR$ of the curvature into one part which only depends on the underlying metric $g$ and another part which represents fluctuations on much shorter distances, is only a first approximation.

A somewhat more realistic model could be obtained by assuming fluctuations on all different length scales:
$$R = R_g + dR_0 + dR_1 + \mathbf{L} + dR_k + \mathbf{L}, \tag{73}$$
where $dR_k$ represents fluctuations on the scale : $l_k = dl^k$ for some appropriate number $l>1$. If we assume all these contributions to be independent, it then follows that we can write
$$\log w_g = \log w_0 + \log w_1 + ... + \log w_k + ..., \tag{74}$$
where now each term, according to Remark 11, can be computed on the corresponding scale, $l_k$. Thus, we get
$$m_g = m_0 + t\|\Re_{g,0}\|^2 + t\|\Re_{g,1}\|^2 \mathbf{L} + t\|\Re_{g,k}\|^2 + \mathbf{L}, \tag{75}$$
where $\Re_{g,k}$ represents the mean Ricci curvature measured on the scale $l_k$. Since the splitting in (73) is in itself an approximation, it is more natural to replace this expression by

$$m_g = m_0 + t \int_d^\infty \frac{c}{k^a} \|\Re_{g,k}\|^2 \, dk, \tag{76}$$

where the constants $c$ and $a$ are related to the parameter $l$ in the definition of the length scales, $l_k$. Thus, we get the following modified expression for $\Xi_G$:

$$\Xi_G = \int_\Omega \left( kr^2 + bR_g^2 + t \int_d^\infty \frac{c}{k^a} \|\Re_{g,k}\|^2 \, dk \right) dV. \tag{77}$$

This formulation has the property of not being local, since the inner integral contains averages of curvature over arbitrarily large sets. This may be interesting from the point of view of quantum mechanics.

We note that the integral in (77) is convergent if $g$ is flat outside some fixed bounded set. On the other hand, we may get serious convergence problems in (77) if $g$ is not asymptotically flat in a sufficiently strong sense. Even more seriously: It is not clear how to make sense of the above formulas in the case when $l_k$ is so large that $g$ can no longer be considered to vary slowly.

In view of this, we will be content with the simpler version Principle (3), especially since the more general Principle (77) is likely to give similar results when applied to gravity.

## 7. Einstein's field equations.

Any theory aiming at an explanation of the concept of mass should include an explanation of Einstein's field equations.

To this end, we simply observe that the integral in Principle 3 is clearly minimized when $R_g = r = \|\Re_g\| = 0$ if there is such a geometry at all. Thus, we get

**Claim 3.** *Consider a region $\Omega$ in space-time with a given finite volume. If there is a geometry which satisfies the boundary conditions such that the Ricci tensor $\Re = 0$ and $r = 0$, then the corresponding macro geometry will maximize the probability.*

*Remark* 13. Clearly, the conclusion is, among other things, based on the approximation that $g$ varies slowly on the scale where the metric fluctuations become important. Thus, the claim can not be used to draw conclusions when we approach the Planck scale. On the other hand, an attempt at explaining Einstein's field equations should also include an answer to why and how they fail to govern reality on the microlevel.

Moreover, it appears that (due to the assumption about fixed volume) Claim 3 should be compatible with some kind of cosmological constant, but this question needs to be further analyzed and we will not go into this in the following.

In free, flat space-time, the conditions $R = 0$, $\Re = 0$ and $r = 0$ are automatically fulfilled. Thus, in this case, the geometric action is equal to zero.

But if we consider a region containing some obstruction, it may be impossible to fulfil these conditions: For instance, the topology may simply not allow a metric with vanishing curvature (Compare with the Gauss-Bonnet theorem ([5]).

However, if the geometry is asymptotically flat, then far away from the obstruction, it is still reasonable to assume that the action is once more minimized by putting $R_{ij} = r = 0$, which in this case (in view of spherical symmetry) leads to the Schwarzschild metric.

We will return to this case in Section 9.

## 8. Particles and Mass-Energy.

In this section, we will try to apply the previous theory to explain the concept of inertial mass of elementary particles. Leaving aside the question about what an elementary particle really is, our basic assumption will be that the presence of a particle manifests itself through a deformation of the topology and geometry of space-time in its vicinity. In accordance with the Principle of Least Action in Section 6, we expect the macro geometries that will actually be realized to be those minimizing $\Xi$ in Principle 3.

Hence, consider a single particle at rest in a space-time where $R_{ij} = 0$. Since, according to the previous section, the action of free space is equal to $0$, we see that the geometric action $\Xi$ of a region only containing one particle, can be associated with the particle itself.

Comparing this to the classical definition of action as proper time times energy (in this case rest energy), we arrive at

**Definition 3.** The *inertial mass* of a particle is defined as
$$M = \Xi/T \tag{78}$$
where $T$ denotes the interval of proper time during which we compute the numerator.

This use of the name inertial mass may seem somewhat arbitrary. However, the following application of the Principle of Least Geometric Action shows that $M$ behaves as inertial mass should with respect to kinetics. For simplicity let us assume the geometry of the surrounding space-time to be the flat one.

**Claim 4.** *Consider a collection of particles in some region in space-time. We assume that they move along straight lines, except that they may momentarily interact by emission and absorption of (real or virtual) particles, thus changing their states of motion and other properties. If we define the total geometric action as the sum of all actions along the world-lines of all particles involved, the Principle of Least Geometric Action implies that the energy of the system, defined as*
$$E = \sum_{i=1}^{n} \frac{M_i}{\sqrt{1 - u_i^2}} \tag{79}$$
*is a conserved quantity, where the $M_i$:s denote the inertial masses of the particles at a given time, and the $u_i$:s denote the corresponding velocities.*

In fact, consider
$$E(t) = \frac{\Delta \Xi}{\Delta t}, \tag{80}$$
where $\Xi$ is the sum of the actions of the individual particles computed in the time-interval $[t, t + \Delta t]$. We now claim that $E(t)$ must be constant as a function of $t$; otherwise we easily construct a volume preserving infinitesimal transformation as in Claim 1, which decreases the action by contracting the time scale at some time $t'$, where $E(t)$ is large and simultaneously expanding the time scale at some other time $t''$ where $E(t)$ is smaller. We conclude that the time-derivate of the total action must be constant. Hence, with

$$\Xi = \sum_{i=1}^{n} M_i T_i = \sum_{i=1}^{n} M_i \sqrt{(t-t_0)^2 - |x_i - x_{0,i}|^2}, \tag{81}$$

we can now compute

$$\frac{\partial}{\partial t}\Xi = \sum_{i=1}^{n} M_i \frac{t-t_0}{\sqrt{(t-t_0)^2 - |x_i - x_{0,i}|^2}} = \sum_{i=1}^{n} \frac{M_i}{\sqrt{1-u_i^2}}, \tag{82}$$

where

$$u_i = \frac{|x_i - x_{0,i}|}{t-t_0}, \tag{83}$$

is the velocity of the $i$:th particle. Thus, the claim follows (see Remark 15 below however).

*Remark* 14. Here, we only consider energy conservation. However, since everything is obviously Lorentz invariant, we must similarly have conservation laws for momentum.

*Remark* 15. It should be noted that the principle of geometric action above is not obviously completely similar to the corresponding usual principle in general relativity, since the particle paths realized there are those *maximizing* the proper time. In fact, it seems likely, but is not evident that a particle generates less action by following a straight line than by following a non-straight path, since the bending of the path itself generates additional curvature. Computer calculations indicate that, for simple worm-hole like topological obstructions, the minimal action is obtained by letting the particle rotate at high speed, which from a macroscopic point of view would seem similar to travelling along a straight line. To the mind of the author, it seems to be a very interesting thought that Newton's first law could depend on a kind of spin property of elementary particles. However, these computations are far from being a proof, and it may very well be that such a proof of the claim (if possible at all), would have to depend on other properties of elementary particles than what is considered here.

Before turning to gravity, let us also note that our postulates give rise to a kind of Heisenberg uncertainty principle. This is perhaps not too surprising since, in a sense, uncertainty was built in from the beginning through Postulates 2 and 4. Nevertheless, it may have some interest to note that the theory gives rise to quantum mechanical effects as well, and that the mass-energy concept that we have introduced behaves as it should in this respect too.

To start with, we note that to measure the mass-energy of a particle in this context means measuring the action integral

$$\Xi_g = \int_\Omega \left( kr^2 + bR_g^2 + t\left\|\Re_g\right\|^2 \right) dV \tag{84}$$

during some interval of time. Due to the fluctuations, however, this measurement will be uncertain with an error $\Delta\Xi$ with probability

$$\exp\{-\Delta\Xi\} \tag{85}$$

according to the theory in Section 4. Hence, the errors with a reasonable chance of occurring must have $\Delta\Xi : 1$.

Now, a measurement of the mass-energy of a particle over an interval of time between $t_1$ and

$t_2$ gives

$$E = \frac{\Xi}{t_2 - t_1}, \qquad (86)$$

where the uncertainty of the numerator is of the order $:1$.

Writing $\Delta T = t_2 - t_1$, we now see that

$$\Delta E : \frac{1}{\Delta T} \qquad (87)$$

which can also be written as

$$\Delta E \Delta T : 1. \qquad (88)$$

Thus, in the units implicit in these calculations, Planck's constant is of the order of unity. Unfortunately, there does not seem to be any easy way of translating this into ordinary units, without making additional assumptions about the nature of mass, particles etc.

## 9. The Schwarzschild mass of a particle.

In Section 8, we have defined the inertial mass of a particle as

$$M = \text{"Geometric action"}/\text{"time"}. \qquad (89)$$

On the other hand, according to the results in Section 7, the metric $g$ of a macro geometry $G$ must (at least far away from topological obstructions) satisfy Einstein's field equations, i.e. the Ricci tensor must vanish, which means that inertial and gravitational mass are equivalent in this sense.

However, there is also another aspect on equivalence. Consider a single particle at rest in space-time. As remarked at the end of Section 7, the geometry far away from the particle must be given by the Schwarzschild metric (with $r = 0$)

$$g = -(1 - \frac{2M_S}{r})dt^2 + (1 - \frac{2M_S}{r})^{-1}dr^2 + r^2 d\varphi^2 + r^2 \sin^2 \varphi dq^2. \qquad (90)$$

Clearly, the constant $M_S$ is determined by the particle; thus we arrive at a third kind of mass which we will refer to as the Schwarzschild mass.

In view of plain physical common sense, it is now very tempting to identify $M_S$ with $M$ above. However, from a mathematical point of view, it is far from obvious that the different concepts of mass are equivalent in this context, and it may in fact very well be that they are not.

In this section, we shall construct a very simplified model for the gravitational field around a particle, where it is possible to see when the action is minimized. In principle, this construction makes it possible to estimate the size of the Schwarzschild mass $M_S$, and thus the strength of the gravitational force.

Hence, let us return to the case of the cylinder $\Omega$ with radius $s$ in Section 4. Now, also introduce a much smaller cylindrical set $B$ with radius $b = s$ at the center of $\Omega$, which represents a particle at rest. In the region $A = \Omega$, $B$, we initially suppose that the geometry is flat. Inside $B$ on the other hand, the geometry of space-time may be distorted and the curvature as well as the density may be non-trivial. To compute the action of the particle, we must integrate the measure

$$d\Xi = \left(kr^2 + bR_g^2 + t\|\Re_g\|^2\right)dV \qquad (91)$$

over $B$. However, for the purpose of estimating $M_S$, we now simply let the geometry within $B$ be flat (with $r = 0$), and replace the curvature terms in (91) by a constant term $xdV$, which represents a kind of average curvature:

$$d\Xi = \left(kr^2 + x\right)dV. \qquad (92)$$

Note that these assumptions imply that the model is in equilibrium with the exterior in the sense of Section 4, when $M_S = 0$.

*Remark* 16. The above model may be considered to represent an "average particle", where all individual properties of the topology have been neglected in favour of a kind of average curvature parameter, $x$. It seems quite reasonable to expect this coarse model to reflect the character of the general case, and to give forces of the right order of magnitude since, in fact, curvature only enters into the action integral in integrated form.

As for the assumption $r = 0$, this is definitely a simplification. Although $r$ is assumed to be small throughout this paper, it can still be expected to interact with the curvature in a complicated way to minimize the action. It would be possible to assign a constant (but non-zero) value to $r$ in $B$. The reader may try to convince himself that, although it complicates the computations below, this assumption would not effect the conclusion in any very important way.

We will now show why the action of this geometry can be reduced by replacing the flat geometry in $A$ by the Schwarzschild metric (90), and also compute the value of $M_S$ which minimizes $\Xi$.

To outline the main idea, we must first ask why the presence of a particle should at all effect the geometry of space-time far away from it. The answer (within the framework of the present theory) is that the curvature terms in $\Xi_G$ should be made as small as possible in order to minimize the action. A step in this direction could be to slow down the flow of time within $B$, exactly as predicted by general relativity. On the other hand, such a contraction will decrease the space-time volume, and as a consequence (as it turns out) decrease the pressure which is in conflict with the equilibrium requirement in Principle 3. The only way out is to assume that the density inside $B$ increases, so as to keep the total pressure constant. The change in action due to the change of time scale is approximately linear in $M_S$, whereas the effect of the change in density turns out to be of second order. Thus, we can expect a minimum for a non-zero $M_S$.

To make this more precise, first note that, in all computations below, the metric within $B$ is unchanged. The only effects taken into account when $M_S$ varies are: (1) The change in time-scale and (2) the change in density within $B$. Furthermore, because of the homogeneity in the time-direction, $A$ and $B$ are unbounded, so for our formulas to make sense, we will carry out all computations per unit of time *at the outer boundary of $A$*.

To find the right density $r_B$ within $B$, we will now imitate the computation from Section 4

for the case of a very weak gravitational field. In this case, it is still reasonable to assume that the modification to the metric when moving the boundary of $\Omega$ the distance $\Delta s$ outwards, is (to first order) linear and homogeneous in the interior of $\Omega$ (in the sense that each cell will change its volume by the same factor).

Under these assumptions, the only choice for the modified metric (to first order and outside of $B$) is

$$g_1 = -(1-\frac{2M_S}{r})dt^2 + l\left[(1-\frac{2M_S}{r})^{-1}dr^2 + r^2\,df^2 + r^2\sin^2 f\,dq^2\right], \qquad (93)$$

where $l$ is now given by

$$l = 1 + (1-\frac{M_S}{s})\frac{\Delta s}{s}. \qquad (94)$$

Note the factor $(1-M_S/s)$ which results from a first order approximation of the radial dilatation of the Schwarzschild metric. Furthermore, note that the coefficient of $dt^2$ is also modified, but only by a second order term which we have omitted. Consequently, (compare (45),

$$\Delta r = (1-\frac{M_S}{s})\frac{3\Delta s}{s}. \qquad (95)$$

Let $V(r)$ denote the volume of a sphere of radius $r$ in Euclidean three-space. It is then easily seen that the volumes of the regions $A$ and $B$ are given by

$$V_A = (1-\frac{2M_S}{s})^{-1/2}(V(s)-V(b)) \approx (1+\frac{M_S}{s})(V(s)-V(b)), \qquad (96)$$

$$V_B = V(b)(1-\frac{2M_S}{b})^{1/2}(1-\frac{2M_S}{s})^{-1/2} \approx V(b)(1-\frac{M_S}{b}), \qquad (97)$$

(per unit of time at $\partial A$). In fact, it is a property of the Schwarzschild geometry that the volume of the region $A = \Omega$, $B$ (per unit of time at infinity) is the same as in the corresponding Euclidean case $= V(s)-V(b)$. Similarly, to get the volume of $B$, (per unit of time at infinity), we must multiply the Euclidean volume $V(b)$ by the time-contraction factor $(1-2M_S/b)^{1/2}$. In both cases, we get the volume per unit of time at $\partial A$ by dividing with $(1-2M_S/s)^{1/2}$. (Note that we have also neglected $2M_S/s$ in comparison with $2M_S/b$ in the second formula, since $b = s$.)

We can now compute the change in action in analogy with (46), but separating the terms corresponding to regions $A$ and $B$ (using (39) and (95)):

$$\Delta \Xi = \Delta \Xi_A + \Delta \Xi_B = p_A \times V_A \times \Delta r + p_B \times V_B \times \Delta r \qquad (98)$$

$$= p_0 \times (1+\frac{M_S}{s})(V(s)-V(b)) \times \left((1-\frac{M_S}{s})\frac{3\Delta s}{s}\right)$$

$$+ (p_0 + 2kr_B) \times V(b)(1-\frac{M_S}{b}) \times \left((1-\frac{M_S}{s})\frac{3\Delta s}{s}\right)$$

$$\approx p_0 4p s^2 \Delta s + 4p b^3 (2kr_B - p_0 \frac{M_S}{b})\Delta s,$$

where we have neglected terms of higher order like $(M_S/s)^2$, $r_B M_S/b$ and also $M_S/s$ in comparison with $M_S/b$. By comparing with (47), we see that to generate the same pressure on the exterior as in the vacuum case, the second term in the last step of (98) must vanish. Thus, we arrive at the following condition for equilibrium for a given $M_S$:

$$r_B = \frac{p_0 M_S}{2kb}. \tag{99}$$

The action of the particle (for a given $M_S$) is given by

$$\Xi = \int_B d\Xi = \int_B \left(kr^2 + x\right) dV = \tag{100}$$

$$= (x + kr_B^2)V_B = (x + kr_B^2)V(b)(1 - \frac{M_S}{b}) \qquad (/\text{u.t.}). \tag{101}$$

In particular, we note that when $M_S = 0$ (and hence also $r_B = 0$), we get

$$\Xi_0 = \int_{B_0} x \, dV = M, \tag{102}$$

where $M$ is the inertial mass of the particle in the approximation of flat space.

We now insert (99) into (100), and neglecting a third-order term containing $(M_S/b)^3$, we get,

$$\Xi \approx \Xi_0 - \Xi_0 \frac{M_S}{b} + k\left(\frac{p_0 M_S}{2kb}\right)^2 V(b). \tag{103}$$

Taking the derivative with respect to $M_S$, we see that the value minimizing the action is given by (with $M = \Xi_0$),

$$M_S = \frac{3k}{16p\, p_0^2 b^2} M. \tag{104}$$

*Remark* 17.   Although the methods that we have used here are by far too crude to allow any predictions in the usual sense of the word, it may still be interesting to note one interesting consequence of formula (104). Contrary to what is commonly supposed, the Schwarzschild mass $M_S$ is here *not* proportional to the corresponding inertial mass. In fact, small particles may have a much larger ratio $M_S/M \propto b^{-2}$ than larger ones. Hence, it might be that the curvature of space-time that we actually observe is mainly generated by leptons, not by heavier particles.

Unfortunately, a difference in the ratio $M_S/M$ between different elementary particles may be difficult to detect directly, since the Schwarzschild mass can only be observed for large quantities of matter, where the number of hadrons and the number of electrons are more or less proportional.

Still, having said this, it is nevertheless interesting to note that there may be an indirect way to observe the difference between inertial and Schwarzschild mass: In fact, if electrons would contribute more to the gravitation that we observe than heavier particles, so should other leptons like neutrinos. Thus, these could be candidates for missing dark matter in the universe.

In fact, when measuring inertial mass, neutrinos are usually not included, whereas we do suppose that they contribute to the curvature of space-time.

Clearly, these questions need to be more deeply investigated before any firm predictions can be made.

Martin Tamm,
Dept of Mathematics,
Univ. of Stockholm,
SE-106 91 Stockholm,
Sweden.
matamm@math.su.se